\documentclass[10pt,a4paper,onecolumn]{article}
\usepackage{marginnote}
\usepackage{graphicx}
\usepackage[rgb]{xcolor}
\usepackage{authblk,etoolbox}
\usepackage{titlesec}
\usepackage{calc}
\usepackage{tikz}
\usepackage[pdfa]{hyperref}
\usepackage{hyperxmp}
\hypersetup{%
    unicode=true,
    pdfapart=3,
    pdfaconformance=B,
    pdftitle={QMKPy: A Python Testbed for the Quadratic Multiple
Knapsack Problem},
    pdfauthor={Karl-Ludwig Besser, Eduard A. Jorswieck},
    pdfpublication={Journal of Open Source Software},
    pdfpublisher={Open Journals},
    pdfissn={2475-9066},
    pdfpubtype={journal},
    pdfvolumenum={},
    pdfissuenum={},
    pdfdoi={10.21105/joss.04718},
    pdfcopyright={Copyright (c) 2022, Karl-Ludwig Besser, Eduard A.
Jorswieck},
    pdflicenseurl={http://creativecommons.org/licenses/by/4.0/},
    colorlinks=true,
    linkcolor=[rgb]{0.0, 0.5, 1.0},
    citecolor=Blue,
    urlcolor=[rgb]{0.0, 0.5, 1.0},
    breaklinks=true
}
\immediate\pdfobj stream attr{/N 3} file{sRGB.icc}
\pdfcatalog{%
  /OutputIntents [
    <<
      /Type /OutputIntent
      /S /GTS_PDFA1
      /DestOutputProfile \the\pdflastobj\space 0 R
      /OutputConditionIdentifier (sRGB)
      /Info (sRGB)
    >>
  ]
}
\usepackage{caption}
\usepackage{orcidlink}
\usepackage{tcolorbox}
\usepackage{amssymb,amsmath}
\usepackage{ifxetex,ifluatex}
\usepackage{seqsplit}
\usepackage{xstring}

\usepackage{float}
\let\origfigure\figure
\let\endorigfigure\endfigure
\renewenvironment{figure}[1][2] {
    \expandafter\origfigure\expandafter[H]
} {
    \endorigfigure
}

\usepackage{fixltx2e} %

\newlength{\cslhangindent}
\newlength{\csllabelwidth}
\newenvironment{CSLReferences}[2] %
 {%
  \setlength{\parindent}{0pt}
  \ifodd #1 \everypar{\setlength{\hangindent}{\cslhangindent}}\ignorespaces\fi
  \ifnum #2 > 0
  \setlength{\parskip}{#2\baselineskip}
  \fi
 }%
 {}
\usepackage{calc}

\usepackage[top=3.5cm, bottom=3cm, right=1.5cm, left=1.0cm,
            headheight=2.2cm, reversemp, includemp, marginparwidth=4.5cm]{geometry}

\titleformat{\section}
  {\normalfont\sffamily\Large\bfseries}
  {}{0pt}{}
\titleformat{\subsection}
  {\normalfont\sffamily\large\bfseries}
  {}{0pt}{}
\titleformat{\subsubsection}
  {\normalfont\sffamily\bfseries}
  {}{0pt}{}
\titleformat*{\paragraph}
  {\sffamily\normalsize}

\usepackage{fancyhdr}
\makeatletter
\let\ps@plain\ps@fancy
\fancyheadoffset[L]{4.5cm}
\fancyfootoffset[L]{4.5cm}

\definecolor{linky}{rgb}{0.0, 0.5, 1.0}

\newtcolorbox{repobox}
   {colback=red, colframe=red!75!black,
     boxrule=0.5pt, arc=2pt, left=6pt, right=6pt, top=3pt, bottom=3pt}

\newcommand{\ExternalLink}{%
   \tikz[x=1.2ex, y=1.2ex, baseline=-0.05ex]{%
       \begin{scope}[x=1ex, y=1ex]
           \clip (-0.1,-0.1)
               --++ (-0, 1.2)
               --++ (0.6, 0)
               --++ (0, -0.6)
               --++ (0.6, 0)
               --++ (0, -1);
           \path[draw,
               line width = 0.5,
               rounded corners=0.5]
               (0,0) rectangle (1,1);
       \end{scope}
       \path[draw, line width = 0.5] (0.5, 0.5)
           -- (1, 1);
       \path[draw, line width = 0.5] (0.6, 1)
           -- (1, 1) -- (1, 0.6);
       }
   }

\patchcmd{\@maketitle}{center}{flushleft}{}{}
\patchcmd{\@maketitle}{center}{flushleft}{}{}
\patchcmd{\@maketitle}{\LARGE}{\LARGE\sffamily}{}{}
\def\maketitle{{%
  
  \AB@maketitle}}
\renewcommand\AB@affilsepx{ \protect\Affilfont}
\renewcommand\AB@affilnote[1]{{\bfseries #1}\hspace{3pt}}
\renewcommand{\affil}[2][]%
   {\newaffiltrue\let\AB@blk@and\AB@pand
      \if\relax#1\relax\def\AB@note{\AB@thenote}\else\def\AB@note{#1}%
        \setcounter{Maxaffil}{0}\fi
        \begingroup
        \let\href=\href@Orig
        \let\protect\@unexpandable@protect
        \def\thanks{\protect\thanks}\def\footnote{\protect\footnote}%
        \@temptokena=\expandafter{\AB@authors}%
        {\def\\{\protect\\\protect\Affilfont}\xdef\AB@temp{#2}}%
         \xdef\AB@authors{\the\@temptokena\AB@las\AB@au@str
         \protect\\[\affilsep]\protect\Affilfont\AB@temp}%
         \gdef\AB@las{}\gdef\AB@au@str{}%
        {\def\\{, \ignorespaces}\xdef\AB@temp{#2}}%
        \@temptokena=\expandafter{\AB@affillist}%
        \xdef\AB@affillist{\the\@temptokena \AB@affilsep
          \AB@affilnote{\AB@note}\protect\Affilfont\AB@temp}%
      \endgroup
       \let\AB@affilsep\AB@affilsepx
}
\makeatother

\renewcommand\Affilfont{\sffamily\small\mdseries}
\usepackage[T1]{fontenc}
\usepackage[utf8]{inputenc}
\IfFileExists{upquote.sty}{\usepackage{upquote}}{}
\IfFileExists{microtype.sty}{%
\usepackage{microtype}
\UseMicrotypeSet[protrusion]{basicmath} %
}{}

\PassOptionsToPackage{usenames,dvipsnames}{color} %
\ifLuaTeX
\usepackage[bidi=basic]{babel}
\else
\usepackage[bidi=default]{babel}
\fi
\babelprovide[main,import]{american}

\def\languageshorthands#1{}

\IfFileExists{parskip.sty}{%
\usepackage{parskip}
}{%
\setlength{\parindent}{0pt}
\setlength{\parskip}{6pt plus 2pt minus 1pt}
}
\providecommand{\tightlist}{%
  \setlength{\itemsep}{0pt}\setlength{\parskip}{0pt}}
\ifx\paragraph\undefined\else
\let\oldparagraph\paragraph
\renewcommand{\paragraph}[1]{\oldparagraph{#1}\mbox{}}
\fi
\ifx\subparagraph\undefined\else
\let\oldsubparagraph\subparagraph
\renewcommand{\subparagraph}[1]{\oldsubparagraph{#1}\mbox{}}
\fi
\ifLuaTeX
  \usepackage{selnolig}  %
\fi

\title{QMKPy: A Python Testbed for the Quadratic Multiple Knapsack
Problem}

\author[1%
]{Karl-Ludwig Besser%
  \,\orcidlink{0000-0002-1597-8963}\,%
}
\author[1%
]{Eduard A. Jorswieck%
  \,\orcidlink{0000-0001-7893-8435}\,%
}

\affil[1]{Institute for Communications Technology, Technische
Universität Braunschweig, Germany}
\date{\vspace{-2.5ex}}

\begin{document}
\maketitle

\marginpar{

  \begin{flushleft}
  \sffamily\small

  {\bfseries DOI:} \href{https://doi.org/10.21105/joss.04718}{\color{linky}{10.21105/joss.04718}}

  \vspace{2mm}

  {\bfseries Software}
  \begin{itemize}
    \setlength\itemsep{0em}
    \item \href{https://github.com/openjournals/joss-reviews/issues/4718}{\color{linky}{Review}} \ExternalLink
    \item \href{https://github.com/klb2/qmkpy}{\color{linky}{Repository}} \ExternalLink
    \item \href{https://doi.org/10.5281/zenodo.7249626}{\color{linky}{Archive}} \ExternalLink
  \end{itemize}

  \vspace{2mm}

  \par\noindent\hrulefill\par

  \vspace{2mm}

  {\bfseries Editor:} \href{https://github.com/jbytecode}{Mehmet Hakan Satman} \ExternalLink
  \,\orcidlink{0000-0002-9402-1982} \\
  \vspace{1mm}
    {\bfseries Reviewers:}
  \begin{itemize}
  \setlength\itemsep{0em}
    \item \href{https://github.com/ulf1}{@ulf1}
    \item \href{https://github.com/max-little}{@max-little}
    \end{itemize}
    \vspace{2mm}

  {\bfseries Submitted:} 08 August 2022\\
  {\bfseries Published:} 02 November 2022

  \vspace{2mm}
  {\bfseries License}\\
  Authors of papers retain copyright and release the work under a Creative Commons Attribution 4.0 International License (\href{https://creativecommons.org/licenses/by/4.0/}{\color{linky}{CC BY 4.0}}).

  \end{flushleft}
}

\hypertarget{summary}{%
\section{Summary}\label{summary}}

QMKPy provides a Python framework for modeling and solving the quadratic
multiple knapsack problem (QMKP). It is primarily aimed at researchers
who develop new solution algorithms for the QMKP. QMKPy therefore mostly
functions as a testbed to quickly implement novel algorithms and compare
their results with existing ones. However, the package also already
includes implementations of established algorithms for those who only
need to solve a QMKP as part of their research.

The QMKP is a type of knapsack problem which has first been analyzed by
Hiley \& Julstrom (\protect\hyperlink{ref-Hiley2006}{2006}). For a basic
overview of other types of knapsack problems see, e.g., Kellerer et al.
(\protect\hyperlink{ref-Kellerer2004}{2004}). As in the regular multiple
knapsack problem, the goal is to assign \(N\in\mathbb{N}\) items with
given weights \(w_i\in\mathbb{R}_{+}\) and (non-negative) profit values
\(p_i\in\mathbb{R}_{+}\) to \(K\in\mathbb{N}\) knapsacks with given
weight capacities \(c_u\in\mathbb{R}_{+}\), such that a total profit is
maximized. In the QMKP, there exist additional joint profits
\(p_{ij}\in\mathbb{R}_{+}\) which are yielded when items \(i\) and \(j\)
are packed into the same knapsack.

Mathematically, the QMKP is described by the following optimization
problem \begin{subequations}
\begin{alignat}{3}
    \max\quad & \sum_{u\in\mathcal{K}}\Bigg(\sum_{i\in\mathcal{A}_u} p_{i} &+&\sum_{\substack{j\in\mathcal{A}_u \\ j\neq i}} p_{ij}\Bigg)\\
    \mathrm{s.t.}\quad & \sum_{i\in\mathcal{A}_u} w_{i} \leq c_u & \quad & \forall u\in\mathcal{K} \\
    & \sum_{u=1}^{K} a_{iu} \leq 1  & & \forall i\in\{1, 2, \dots{}, N\}
\end{alignat}
\end{subequations} where \(\mathcal{K}=\{1, 2, \dots{}, K\}\) describes
the set of knapsacks, \(\mathcal{A}_u\subseteq\{1, 2, \dots{}, N\}\) is
the set of items that are assigned to knapsack \(u\) and
\(a_{iu}\in\{0,1\}\) is a binary variable indicating whether item \(i\)
is assigned to knapsack \(u\).

\hypertarget{statement-of-need}{%
\section{Statement of need}\label{statement-of-need}}

The QMKP is an NP-hard optimization problem and therefore, there exists
a variety of (heuristic) algorithms to find good solutions for it. While
Python frameworks already exist for the standard (multiple) knapsack
problem and the quadratic knapsack problem, they do not consider the
\emph{quadratic multiple} knapsack problem. However, this type of
problem arises in many areas of research. In addition to the typical
problems in Operations Research, it also occurs in distributed computing
(\protect\hyperlink{ref-Rust2020}{Rust et al., 2020}) and in the area of
wireless communications (\protect\hyperlink{ref-Besser2022wiopt}{Besser
et al., 2022}).

For the classic knapsack problem and the quadratic \emph{single}
knapsack problem, many well-known optimization frameworks like Gurobi
(\protect\hyperlink{ref-gurobi}{Gurobi Optimization, LLC, 2022}) and
OR-Tools (\protect\hyperlink{ref-ortools}{Perron \& Furnon, 2022})
provide solution routines. However, they are not directly applicable to
the QMKP. Furthermore, it can be difficult for researchers to reproduce
results that rely on commerical software.

Therefore, QMKPy aims to close that gap by providing an open source
testbed to easily implement and compare solution algorithms.
Additionally, Python is widely used among researchers and enables easy
to read implementations. This further supports the goal of QMKPy to
promote sharable and reproducible solution algorithms.

For initial comparisons, the software already implements multiple
solution algorithms for the QMKP, including a \emph{constructive
procedure (CP)} based on Algorithm 1 from Aïder et al.
(\protect\hyperlink{ref-Aider2022}{2022}) and the greedy heuristic from
Hiley \& Julstrom (\protect\hyperlink{ref-Hiley2006}{2006}). A second
algorithm that is included is the \emph{fix and complete solution (FCS)
procedure} from Algorithm 2 in Aïder et al.
(\protect\hyperlink{ref-Aider2022}{2022}). Additionally, a collection of
reference QMKP instances that can be used with QMKPy is provided at
(\protect\hyperlink{ref-QMKPyDatasets}{Besser, 2022}). This dataset
includes the well-known reference problems from Hiley \& Julstrom
(\protect\hyperlink{ref-Hiley2006}{2006}), which in turn are based on
the quadratic single knapsack problems from Billionnet \& Soutif
(\protect\hyperlink{ref-Billionnet2004}{2004}).

The open source nature of QMKPy and its aim at researchers encourages
the implementation of more algorithms for solving the QMKP that can
become part of the QMKPy framework.

The most notable benefits when implementing an algorithm using QMKPy are
the following:

\begin{itemize}
\tightlist
\item
  No additional overhead is required. Only a single function with the
  novel solution algorithm needs to be implemented.
\item
  Generic unit tests are available to make sure that the novel algorithm
  fulfills the set of basic criteria.
\item
  The ability of loading and saving problem instances allows for quick
  and easy testing of any algorithm against reference datasets. This
  enables reproducible research and creates a high degree of
  comparability between different algorithms.
\end{itemize}

\hypertarget{references}{%
\section*{References}\label{references}}
\addcontentsline{toc}{section}{References}

\hypertarget{refs}{}
\begin{CSLReferences}{1}{0}
\leavevmode\vadjust pre{\hypertarget{ref-Aider2022}{}}%
Aïder, M., Gacem, O., \& Hifi, M. (2022). Branch and solve
strategies-based algorithm for the quadratic multiple knapsack problem.
\emph{Journal of the Operational Research Society}, \emph{73}(3),
540–557. \url{https://doi.org/10.1080/01605682.2020.1843982}

\leavevmode\vadjust pre{\hypertarget{ref-QMKPyDatasets}{}}%
Besser, K.-L. (2022). \emph{QMKPy datasets} (Version v1.0) {[}Computer
software{]}. Zenodo. \url{https://doi.org/10.5281/ZENODO.7157144}

\leavevmode\vadjust pre{\hypertarget{ref-Besser2022wiopt}{}}%
Besser, K.-L., Jorswieck, E. A., \& Coon, J. P. (2022, September).
Multi-user frequency assignment for ultra-reliable {mmWave} two-ray
channels. \emph{20th International Symposium on Modeling and
Optimization in Mobile, Ad Hoc, and Wireless Networks (WiOpt)}.

\leavevmode\vadjust pre{\hypertarget{ref-Billionnet2004}{}}%
Billionnet, A., \& Soutif, Éric. (2004). Using a mixed integer
programming tool for solving the 0–1 quadratic knapsack problem.
\emph{INFORMS Journal on Computing}, \emph{16}(2), 188–197.
\url{https://doi.org/10.1287/ijoc.1030.0029}

\leavevmode\vadjust pre{\hypertarget{ref-gurobi}{}}%
Gurobi Optimization, LLC. (2022). \emph{{Gurobi Optimizer Reference
Manual}}. \url{https://www.gurobi.com}

\leavevmode\vadjust pre{\hypertarget{ref-Hiley2006}{}}%
Hiley, A., \& Julstrom, B. A. (2006). The quadratic multiple knapsack
problem and three heuristic approaches to it. \emph{Proceedings of the
8th Annual Conference on Genetic and Evolutionary Computation - GECCO
’06}, 547–552. \url{https://doi.org/10.1145/1143997.1144096}

\leavevmode\vadjust pre{\hypertarget{ref-Kellerer2004}{}}%
Kellerer, H., Pferschy, U., \& Pisinger, D. (2004). \emph{Knapsack
problems}. Springer Berlin Heidelberg.
\url{https://doi.org/10.1007/978-3-540-24777-7}

\leavevmode\vadjust pre{\hypertarget{ref-ortools}{}}%
Perron, L., \& Furnon, V. (2022). \emph{OR-tools} (Version v9.4)
{[}Computer software{]}. Google.
\url{https://developers.google.com/optimization/}

\leavevmode\vadjust pre{\hypertarget{ref-Rust2020}{}}%
Rust, P., Picard, G., \& Ramparany, F. (2020). Resilient distributed
constraint optimization in physical multi-agent systems. \emph{24th
European Conference on Artificial Intelligence (ECAI 2020)}, 195–202.
\url{https://doi.org/10.3233/FAIA200093}

\end{CSLReferences}

\end{document}